\input{aipcheck}

\documentclass[
    ,final            
  ]
  {aipproc}

\layoutstyle{6x9}

\begin{document}

\title{Early metal enrichment in high-redshift quasars}

\classification{98.54.Aj - 98.58.Ay - 98.62.Bj - 98.62.Js}
\keywords      { galaxies: abundances,
 galaxies: high-redshift, quasars: emission lines}

\author{R.~Maiolino}{address={Astronomical Observatory of Rome, via di Frascati 33,
  00040 Monte Porzio Catone, Italy}
}

\begin{abstract}
Quasars are powerful systems whose spectrum is rich of metal features
that allow us to investigate the chemical evolution of galaxies
at very high redshift, even close to the reionization epoch.
I review the main observational constraints on the
metallicity of quasars host galaxies at high redshift and discuss
the implications and issues for models of galaxy evolution in the
early universe.
\end{abstract}

\maketitle


\section{Introduction}

Thanks to their huge luminosity, quasars deliver high signal-to-noise spectra
even at very high redshift. The most distant quasars currently known are
at z$\sim$6.4, at the end of the re-ionization epoch \citep{fan06,willott07}.
The several metal features observed in their spectrum 
provide precious information on the metal enrichment of
their host galaxies, which is linked to their past star formation history.
Luminous quasars and radio galaxies are generally hosted in very massive
systems, therefore
they offer the possibility to investigate the metallicity evolution
of the most massive representatives
of the galaxy population at any epoch.
Generally, metallicity information on quasars is inferred from the analysis of the
broad emission lines, which are very strong, but are emitted from a
very small nuclear region. Metallicity information on the
scales of the host galaxy can be inferred from the narrow emission
lines, which are emitted from a region ranging from a few 100 pc to several
kpc in size; however
narrow lines at high-z are generally detectable only in those
systems where the broad lines are obscured, i.e. type 2 quasars
and narrow line radio galaxies.
An alternative method to investigate the metallicity of quasars host galaxies
is to measure metallicity tracers emitted from the host galaxy, not
excited by the nuclear accreting black hole, but powered by star formation;
however these tracers are generally observable in the millimeter
and far-IR spectral ranges, where observations are still limited to a
few, extremely bright objects. Each of these metallicity measurements provides
precious and complementary
information on the evolutionary stage and star formation history
of quasar host galaxies in
the early universe.
In this paper I review these various
observational results on the metallicity enrichment of quasars and
radio galaxies, focusing on the most distant systems known, and I review some
important implications of these observational findings
for our understanding of
galaxy formation in the early universe.

\section{Observational constraints on the metallicity of high-z AGN}

\subsection{The metallicity of the BLR in high-z quasars}

Thanks to the large spectroscopic SDSS quasar sample, and by
exploiting complex photoionization models, it has been possible
to constrain the metallicity of the Broad Line Region (BLR, i.e.
the innermost ionized region)
in large samples of quasars in the
redshift range 2$<$z$<$4.5 (which is the redshift range where
the main UV broad lines, used to constrain the BLR metallicity, are
observable in the optical) \citep{hamann99,nagao06a}.
In most cases the BLR metallicity is inferred to be several times solar.
Moreover, there is a clear dependence of
the BLR metallicity on the quasar luminosity. The metallicity-luminosity
relation may actually result from a black hole-metallicity dependence
\citep{warner04} or from a relation between accretion rate (L/L$_{Edd}$)
and metallicity \citep{shemmer04}. However, an interesting result is that,
once the metallicity-luminosity relation is accounted
for, there is no evidence for any evolution of the BLR metallicity with
redshift \cite{nagao06a}.
At redshift z$>$5, i.e. approaching the re-ionization epoch, determining the
metallicity of the BLR is more difficult, since several of the emission lines
used to constrain the metallicity are shifted into the near-IR.
Early near-IR spectroscopic studies found evidence for
very high metallicities of the BLR in a few z$\sim$6 quasars
\cite{pentericci02,jiang07}.
More recently, Juarez et al. \cite{juarez09} used near-IR and
optical spectra to constrain the metallicity of
30 quasars in the redshift range 4$<$z$<$6.4.
They confirm very high metallicities and no
evidence for any metallicity evolution when compared with lower redshift
quasars (Fig.1).

\begin{figure}
  \includegraphics[height=.3\textheight]{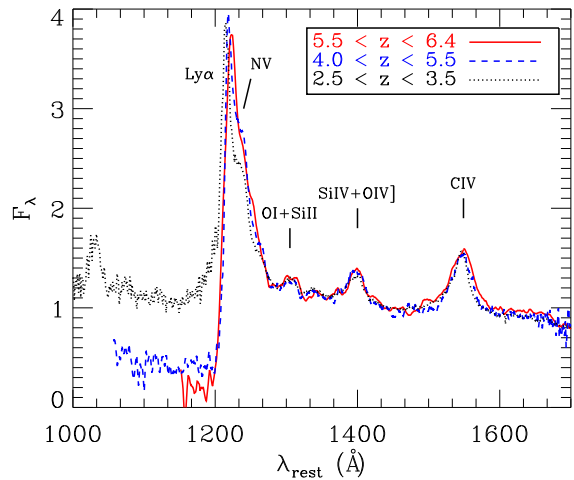}
  \caption{Stacked spectra of quasars in different redshift bins (from Juarez et al.
  \cite{juarez09}). Note that the relative intensity of the metal lines (and in
  particular the (SiIV+OIV])/CIV ratio) remains constant over the wide redshift
  interval 2.5$<$z$<$6.4, indicating
  that the metallicity in the observed quasars 
  does not evolve with redshift.}
\end{figure}

The broad emission lines can also provide clues on relative
chemical abundances. In particular, the abundance of iron and carbon,
which are subject to a delayed enrichment relative to $\alpha$-elements,
provides important information on the star formation history
in quasar hosts.
The UV spectrum of quasars is
characterized by a prominent Fe-bump, around 2300--3100~\AA , due to the blend
of several thousand FeII transitions, whose ratio with the MgII$\lambda$2798 doublet
is used to trace the Fe/$\alpha$ abundance ratio. However, inferring an absolute
value of Fe/$\alpha$ from the FeII/MgII intensity ratio is very difficult, since
the latter depends also on the physical conditions of the BLR \citep{verner04}.
As a consequence, the FeII/MgII is often take as a tracer
of the {\it relative} variations of Fe/$\alpha$.
Several studies have been published
reporting measurements of FeII/MgII
in high redshift quasars (3$<$z$<$6.4), by using deep near-IR spectroscopic
observations \citep{freudling03,iwamuro04,maiolino03,kurk07,dietrich03,sameshima09}.
All of these studies
consistently find that there is no evolution of the FeII/MgII ratio
as a function of redshift,
out to z$=$6.4.
A similar result is obtained for the carbon abundance: the intensity of
CIV$\lambda$1549 and CIII]$\lambda$1900 relative to other lines suggest
that the abundance of carbon relative to other elements remains constant
out to z$=$6.4. I will discuss the implications of these findings in the
next section.

\subsection{Metallicity in the NLR of high-z AGNs}

The BLR is a very small region which may not be representative of the metallicity in
the host galaxy. Gas on large scales,
in the host galaxy, ionized by the AGN produces the so called
Narrow Line Region (NLR).
These narrow lines can generally be measured more easily 
if the nuclear region is
obscured, i.e. in type 2 AGNs. However, in high redshift objects (mostly
radio-galaxies and type 2 QSOs) there are often only a few, faint narrow lines detected
that can be used to constrain the NLR metallicity \citep{debreuck00,
vernet01,iwamuro03,nagao06b}. As a consequence of the faintness of the narrow lines,
the sample of high-z objects for which the NLR metallicity has been investigated
is much smaller ($\sim$60 objects) than the broad line sample, and currently
limited to z$<$4.
By exploiting photoionization models to interpret the narrow line
ratios, it is found that the metallicity of the NLR is generally lower than
typically observed in the BLR, but still super-solar. On average
there is no evidence for any evolution with redshift of the NLR metallicity. 
It is also interesting to note that the lack of evolution observed in the NLR
applies also to the CIV$\lambda$1549/HeII$\lambda$1640 narrow line ratio, suggesting
a lack of evolution of the carbon abundance, as for the BLR, but on much larger
scales.

\subsection{Abundances from the associated absorption systems}

The resonant (UV) narrow absorption lines due to the interstellar
medium of quasars host galaxies provide a powerful tool to investigate the
chemical abundances. When detected, these absorption systems can in
principle provide very accurate measurements of the abundances of individual
elements, since we can directly measure the column of each element (provided
that the absorption lines are not heavily saturated and that the ionization
corrections can be inferred). In practice, this approach
requires high spectral resolution and high S/N spectra, resulting in a small
number of quasars where this strategy can be pursued, most of which at z$<$3.
Generally quasars associated absorption systems show super-solar metallicities,
even for what concerns the carbon abundance (see \cite{dodorico04,gabel06} and
references therein), supporting the findings obtained for the NLR with higher
precision. However, it is important to bear in mind two caveats.
The location of the absorber is unknown; we only know that
it cannot be too close to the nuclear source based on the line width and also based
on the ionization structure (in some cases inferred lower limits on the distance
from the nuclear source are $>$100 ~kpc). As a consequence we do not really know
whether the gas being probed in absorption is part of the quasar host galaxy,
or gas clouds in their halo or in the galaxy outskirts. For what concerns the carbon
abundance, it is important to bear in mind that dust depletion reduces the 
column of carbon measured from the absorption systems. Hence, the carbon abundance
inferred with this method is actually a lower limit.


\subsection{Metallicity constraints from (sub-)mm observations}

Recent mm and submm observations have delivered precious information on the
interstellar medium of high-z quasar host galaxies. The detection of strong
CO emission in some quasars at z$>$5 \citep{walter03,carilli07,maiolino07a},
as well as the detection of the [CII]158$\mu$m
line \citep{maiolino05,iono06,walter09}, with intensities comparable
to lower redshift and local galaxies (with similar star forming rates),
strongly suggests that the host galaxy of these distant quasars is highly
metal enriched, and especially in terms of carbon abundance. These data are still
far from providing accurate measurements of the metallicity and of the chemical
abundances, however future facilities (e.g. ALMA, SPICA)
will allow us to measure additional
transitions (such as [OI]63$\mu$m, [NII]122$\mu$m and other molecular transitions)
that will provide constraints the metal abundances in the ISM \citep{maiolino08b}.

Another indication of high metal abundances in distant
quasars is the detection of huge of dust mass masses (exceeding $\rm 10^8~M_{\odot}$),
as inferred from the mm-submm continuum (far-IR rest-frame) even in the
most distant quasars known, close to the re-ionization epoch
\citep{beelen06,wang07}. Inferring the metallicity from the
dust content is difficult at high redshift
for various reasons: the dust composition is not
well known \citep{maiolino04,stratta07,
willott07}; it is not clear what fraction of the metals have been locked in dust;
finally, both dust mass and gas mass (when available) are subject to
large uncertainties. However,
rough estimates suggest that the observed dust masses must be associated 
to an ISM with solar or super-solar metallicity. If confirmed with more accurate
measurements, these findings provide further evidence that even the host
galaxies of quasars close to re-ionization (z$\sim$6) have reached high metallicities
in these early epochs in the universe,
confirming the results obtained for the BLR, but on much larger scales.

\section{Interpretation and implications for galaxy evolution}

\subsection{The apparent lack of metallicity evolution}

The apparent lack of redshift evolution of the BLR metallicity 
may be deceiving. It does not imply that the metallicity of the BLR in individual
AGNs does not evolve with time. It simply means that quasars that are
detectable (with a given luminosity) at any epoch
have reached the same metallicity regardless
of their redshift. This effect probably results from a combination of selection effects
and the evolutionary link between supermassive black hole and their host galaxies.
Indeed, in a flux-limited survey, quasars can only be detected if they reach a
minimum luminosity, which translates into a minimum black hole mass (even if they
are accreting at the Eddington rate). Most models predict that the quasar host galaxy
evolve along with the nuclear black hole
\citep{granato04,dimatteo05,hopkins08,menci08,li07}, so to match the local $\rm M_{BH}-M_{gal}$
relation. As a consequence, by the time a quasar
has reached the luminosity to be detected, its host galaxy has already evolved and
enriched the interstellar medium. To illustrate this effect more clearly,
Fig.~2 (from \cite{juarez09})
shows the evolution of the metallicity, of the quasar luminosity and of the
gas mass according to one of the theoretical models \cite{granato04}.
The shaded area indicates the epoch when the high-z quasar is luminous enough
to be detected in the SDSS survey; by the time the quasar has reached this limit
the host galaxy has already reached a metallicity of three times solar.
Obscuration may be another parameter introducing additional selection effects.
Models expect that during the early phases the quasar is heavily embedded in gas and dust,
making it unobservable at optical wavelengths. Only during the latest stages the
quasar wind is powerful enough to expel large quantities of dust and gas, making
the quasar detectable by optical surveys. The hatched region in Fig.~2 shows the
epoch when the quasar is unobscured (when
more than half of the gas mass has been expelled), corresponding to an epoch
when the gas metallicity has reached $\rm \sim 4~Z_{\odot}$.

\begin{figure}
  \includegraphics[height=.57\textheight]{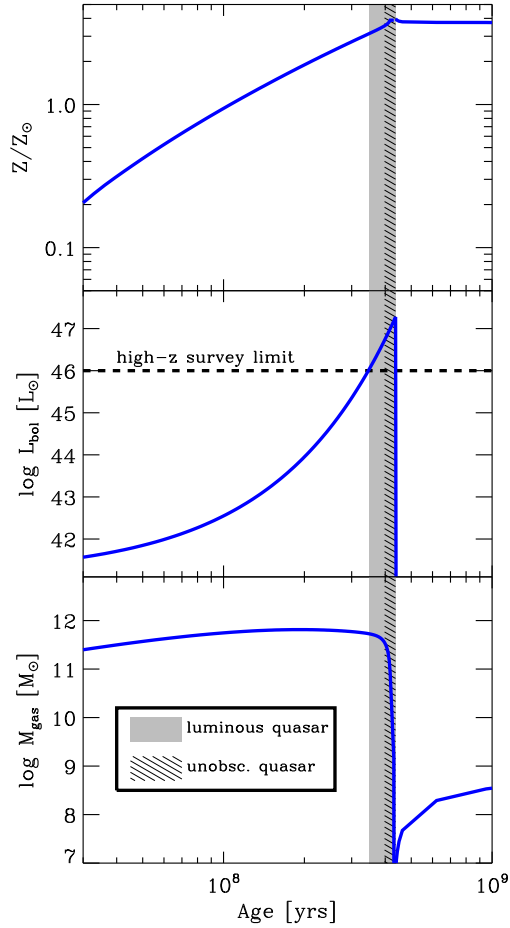}
  \caption{Evolution of the gas metallicity, bolometric quasar luminosity and gas mass
according to the model in \cite{granato04}.
The horizontal dashed line shows the approximate minimum bolometric luminosity detectable at high-z in the SDSS quasar survey.
The shaded area shows the epoch when a high-z quasar is luminous enough to be detected in the SDSS survey (if unobscured).
The hatched area is the epoch when the quasar is likely unobscured and
therefore detectable in optical surveys. By the time the quasar has reached the
luminous, ``unobscured'' epoch the
ISM is already highly enriched. From Juarez et al. \cite{juarez09}.}
\end{figure}

The finding that the BLR metallicities of quasars observed at different redshifts
have on average the same metallicity implies that the evolutionary BH-galaxy
connection occurs in a similar manner at any epoch, regardless of redshift.
In particular, the finding that even in z$\sim$6 quasars the metallicity is 
similar to lower redshift quasars implies that the same BH-galaxy co-evolutionary
mechanism is already at work right after re-ionization.
The very high metallicity observed at z$\sim$6 indicates that the
host galaxies of these quasars are already highly evolved (at least from the chemical
point of view) soon after re-ionization.
These objects probably represent the most extreme cases of galaxy downsizing (at
least in its chemical version
\cite{maiolino08a}), i.e.
these are the most extreme examples of massive galaxies forming very rapidly at very high
redshift.

Note that the same reasoning can explain the apparent
lack of metallicity evolution for the NLR of radio-galaxies
and obscured quasars. Obviously, for these classes of AGNs the selection effects due
to obscuration are unlikely to play a significant role. However, the requirement
of a minimum power to be detected in radio or X-ray surveys (hence minimum
black hole mass) probably produces the same selection effect in terms
of NLR metallicity as for the BLR.

\subsection{The extreme metallicities in the BLR}

According to the photoionization models, the metallicity in quasar nuclei,
as inferred from the broad lines, is extremely high: $\rm Z_{BLR}\sim 7~Z_{\odot}$
for the most distant quasars \citep{juarez09} and in some cases 
approaching $\rm Z_{BLR}\sim 10~Z_{\odot}$. These metallicities are
not unrealistic. Indeed, the BLR is a very small
region, containing only $\sim \rm 10^4~M_{\odot}$ of gas; therefore, even
a few SNe can rapidly enrich the BLR to super-solar metallicities
(more specifically, a SN rate of $\rm 10^{-4} ~yr^{-1}$ can enrich the
BLR to Z$>$Z$_{\odot}$ in less than 10$^8$ years). More puzzling is that
such huge metallicities are not found in the stellar population of massive
galaxies in the local universe. This implies that the high metallicity BLR gas
must be either expelled out of the galaxy, or mixed with lower metallicity gas
in the host galaxy before forming stars.

\subsection{The iron and carbon problems}

Iron is
mostly produced by SNIa, most of which are expected to explode on timescales longer
than $\sim$1~Gyr. Hence, the abundance of iron relative to $\alpha$-elements
(e.g. O, Ne, Mg, which are promptly produced by SNII on very short timescales) is
regarded as a ``clock'' of star formation. The lack of evolution of the
the FeII/MgII ratio as a function of redshift, suggests a lack of Fe/$\alpha$
evolution
out to z$=$6.4. This result is puzzling, since at z$>$5 the age of the universe is
less than 1~Gyr, hence stellar evolution should fall short of time to produce SNIa.
Some possible scenarios can explain this paradox. First, the UV iron bump may not
be a good tracer of iron abundance at all. Indeed the UV Fe bump is a strong coolant of
the BLR and, therefore, its intensity may remain constant independently of the iron
abundance. A better tracer of the iron abundance is probably the optical Fe bump
at $\rm \lambda \sim 4500\AA$ \citep{verner04,netzer07}, but which cannot be measured
from ground at z$>$4, so this test must await JWST. Another possibility is that
the assumed iron enrichment timescale is wrong. Mannucci et al.
\cite{mannucci05,mannucci06} show evidence for a significant
population of SNIa, that explode on timescales as short as 10$^8$~yr.
However, Matteucci et al. \cite{matteucci06}
shows that these ``prompt SNIa'' represent only
$\sim$30\% of the total population integrated over long time scales, hence
this scenario may not provide a viable solution.

A similar timescale problem applies to carbon. Carbon is
mostly produced by AGB stars and planetary nebulae. Although the first AGBs are produced
as soon as 50~Myr after the onset of star formation, most of the carbon production
occurs after $\sim$1~Gyr, resulting into a delayed carbon enrichment relative to
$\alpha$-elements. The finding that the ratio (SiIV+OIV])/CIV does not evolve with
redshift, out to z=6.4, suggests that the abundance of carbon relative to the
$\alpha$-elements does not evolve, contrary to the expectation that the
carbon abundance should drop at z$>$5 because stellar evolutionary timescales fall
short of time to produce carbon. This ``carbon problem'' is not only found in the BLR,
but it is probably an issue arising also from carbon tracers on larger scales in the
host galaxy (carbon lines from the narrow line region, as well as [CII] and CO emission
from the host galaxy).
Both iron and carbon issues at z$\sim$6 remain open and certainly
require additional observations and modelling to be tackled.

\subsection{Quasar metallicity and stellar mass}

The high metal abundances in quasars at high redshift (especially at z$\sim$6)
require that the
host galaxy has undergone a rapid and powerful star formation event. Infrared
and mm observations have indeed revealed vigorous star formation in the host
galaxies of distant quasars, although not ubiquitously
\citep{beelen06,wang07,lutz07,lutz08,maiolino07b}. However, a potential problem
is the comparison of the high level of metal enrichment with the estimated
galaxy mass. Indeed, there are indications that high-z quasars are characterized
by a galaxy--to--black hole mass ratio
at least one order of magnitude lower than observed
in local galaxies \citep{walter04,maiolino07a}.
While the ISM has already reached supersolar metallicities,
more than 90\% of the final stellar mass has still to be formed (to reach
the local $\rm M_{BH}/M_{galaxy}$ relation). If confirmed with additional
observations on a larger sample of quasars, these results may be challenging to
explain by models of early BH-galaxy coevolution.





\bibliographystyle{aipproc}   

\end{document}